# Automating Sonologists' USG Commands with AI and Voice Interface

Emad Mohamed, Shruti Tiwari, Sheena Christabel Pravin*
School of Electronics Engineering,
Vellore Institute of Technology, Chennai.
*sheenachristabel.p@vit.ac.in

**Abstract**

This research presents an advanced AI-powered ultrasound imaging system that incorporates real-time image processing, organ tracking, and voice commands to enhance the efficiency and accuracy of diagnoses in clinical practice. Traditional ultrasound diagnostics often require significant time and introduce a degree of subjectivity due to user interaction. The goal of this innovative solution is to provide Sonologists with a more predictable and productive imaging procedure utilizing artificial intelligence, computer vision, and voice technology. The system's functionality employs computer vision and deep learning algorithms, specifically adopting the Mask R-CNN model from Detectron2 for semantic segmentation of organs and key landmarks. This automation improves diagnostic accuracy by enabling the extraction of valuable information with minimal human input. Additionally, it includes a voice recognition feature that allows for hands-free operation, enabling users to control the system with commands such as "freeze" or "liver," all while maintaining their focus on the patient. The architecture comprises video processing and real-time segmentation modules that prepare the system to perform essential imaging functions, such as freezing and zooming in on frames. The liver histopathology module, optimized for detecting fibrosis, achieved an impressive accuracy of 98.6%. Furthermore, the organ segmentation module produces output confidence levels between 50% and 95%, demonstrating its efficacy in organ detection. In conclusion, the integration of voice commands, organ detection, and a comprehensive analysis of liver histopathology in this project ensures a holistic approach to ultrasoundimaging. The streamlined functionality promises early pathology detection, faster diagnostics, and aligns with the practical needs of dynamic healthcare settings, ensuring a user-friendly experience for clinicians.

**Keywords:** USG, CNN, KNN, AI, NLP, API, COCO, JSON

# 1. Introduction

In modern healthcare, diagnostic imaging plays a pivotal role in identifying, diagnosing, and monitoring a wide array of conditions. Ultrasound imaging is one of the most common diagnostic methods used across nearly every medical department today. Its popularity stems from its real-time imaging capability, non-invasive nature, and easy accessibility in many clinics. This imaging technique helps in examining the internal anatomy of patients, making it valuable for assessing, managing, and, most importantly, diagnosing various diseases, especially in obstetric, cardiovascular, and liver conditions.

The reliance on screens for immediate feedback allows clinicians to make quick decisions regarding both minor and major procedures. However, these advantages are not always apparent with traditional ultrasound systems, which can still be challenging for sonologists. Image interpretation relies on human skill, which can vary significantly, potentially affecting the consistency and accuracy of diagnoses. Additionally, excessive manual control and analysis can lead to physician burnout, particularly during lengthy or complex procedures. Manual adjustments and repetitive tasks during ultrasound procedures may also potentially impact diagnostic consistency and can lead to delayed decision-making. These concerns highlight the urgent need for ultrasound systems that can reduce the burden of manual tasks, enhance diagnostic accuracy, and ensure a smooth workflow.

This project aims to address some of these issues by designing a modern ultrasound machine integrated with AI technology, featuring numerous automations to simplify the diagnostic process. By leveraging computer vision and deep learning, the proposed system automates essential aspects of ultrasound imaging, including organ detection, segmentation, and liver histopathological analysis, while also offering hands-free operational control through voice commands. The objectives of the project are:

• **Automated Control of Ultrasound Devices Using Voice Commands**: This system utilizes a speech recognition module that allows sonologists to perform hands-free functions such as freezing, zooming, or highlighting the liver. By minimizing physical contact with the ultrasound device, caregivers can focus more on patient interaction, ensuring an uninterrupted workflow.

• **Organ Detection and Labeling with AI and Curvature Correction**: The system employs deep learning via the Mask R-CNN model from Detectron2 to detect and annotate important body organs, such as the liver and kidneys. It accurately accommodates the shape of each organ, enabling precise marking of regions. This capability enhances recognition and segmentation accuracy, reducing the need for manual adjustments while increasing the reliability and efficiency of the diagnostic process.

• **Liver Histopathology Analysis**: The system includes specialized modules for evaluating liver histopathology, specifically aimed at detecting liver fibrosis and other pathological changes. After training, these modules demonstrated outstanding performance, thus allowing physicians to timely and accurately identify liver lesions.

By integrating voice control features, automated organ segmentation, and liver histopathology analysis, this system presents a ground breaking ultrasound imaging solution. It provides clinicians with an interactive and intelligent tool that is efficient and precisely equipped with a user-friendly interface, ultimately leading to improved patient care and supporting healthcare practitioners in challenging clinical environments.

## 2. Literature Survey

### 1. Automatic Recognition of Abdominal Organs in Ultrasound Images

The review work in this article is geared towards the application of deep learning techniques for recognizing abdominal organs automatically in ultrasound images. The authors used ResNet and DenseNet along with KNN. Their method uses feature vector-based techniques on the ultrasound images instead of mere distance metrics to findseveral abdominal organs very precisely. It is observed that the deep learning models appear to be useful in organ recognition where good classification results are obtained and it is more so in the instances where different types of organs are presented in the images. Nevertheless, the method calls for excessive amounts of computational power in training the models, which may affect its feasibility in practical situations where resources are limited.

### 2. Towards Clinical Application of AI in Ultrasound Imaging

The potential uses of artificial intelligence (AI) in ultrasound imaging in the clinical field are assessed in this paper through a review of recent developments in deep learning models for pre-processing, segmentation and detection tasks. The authors underline the potential of AI to increase the accuracy and speed of diagnostic processes. In addition, they consider the issue of AI explainability in healthcare, which can help build the confidence of doctors in the use of such systems to a considerable degree. There are many advances that are promising in theory, but practical application in real life is restricted by issues such as integration with existing healthcare systems and compliance with regulations, the authors point out.

### 3. Voice-Controlled Medical Ultrasound Systems

This paper proposes a novel ultrasound imaging and scanning system which can be operated using voice control and aims to optimize medical imaging work flow. The system includes speech recognition, thus, it is possible to operate the system without hands, in which case a clinician can assign tags to the images and control the ultrasound device during scans without using any part of the body to touch the system. This mitigates human error in tagging images and hastens the scan process. Nevertheless, the authors also point out the current speech recognition systems are limited in their application, particularly in environments with excess noise, and that more developments are required to make them effective across varying clinical environments.

### 4. Automated Measurement and Calcification Detectionin Carotid Ultrasound

The present work describes a fully automated method to detect and quantify calcifications in carotid ultrasound images, employing advanced image segmentation techniques along with deep learning models. This system improves the diagnostic care of vascular calcifications, prompting a significant emphasis on their inclusion, which are important cardiovascular risk factors. The authors argue that the proposed approach is capable of improving the accuracy of the diagnosis which, in turn, translates into better treatment results. Nevertheless, the system may fail in its intended purpose if, for example, low-quality sonograms or a high level of sound interference exists which hinders the use of the system in certain clinical situations. Still, such an approach is an asset when it comes to the evaluation of cardiovascular status in a non-invasive manner.

## 5. Deep Learning-Based Tumor Detection in Breast Ultrasound Imaging

The research concern of Johnson et al. finds application of advanced computing techniques such as deep convolution neural networks in real-time detection of tumours in breast ultrasounds. Using the CNN architecture, the authors illustrate that the model can effectively find tumors, hence contributing to the early breast cancer diagnosis dilemma. The importance of this real-time detection feature cannot be overemphasized, especially in combating the problem of delays in diagnosis for better treatment outcome. It is however noted by the writers of the article that in some occasions the constructs may be hyper sensitive to the stimuli and hence generates false positive results when for example a harmless structure is incorrectly regarded as a tumour and this shamelessly points to the need for further improvements and tests of the model in that respect.

## 6. AI-Based Liver Segmentation in Ultrasound Imaging

The research described in this paper looks at the use of AI technology in liver segmentationtasks from ultrasound images with a specific focus on the use of convolutional neural networks (CNN) for the automated and accurate delineation of liver boundaries. The authors state that the method supported with AI reduces the need for human operations which enhances the efficiency and evenness of clinical processes within the laboratories. Their solution also controls the variability of segmentation results that is advantageous in most of the multicenter studies. At the same time, the authors point out that the technique works well only in confined parts of the anatomy, while the type of input training data is critical to the success of segmentation, which is why the technique isnot applicable in a wide range of situations.

## 7. Integrated Voice and Image Processing in Ultrasound Diagnostics

The authors of this manuscript present a machine learning model that combines speech recognition and image analysis to improve the efficiency of ultrasound diagnosis. The purpose of the system is to help the clinicians arrive at faster and more accurate diagnostic conclusions by eliminating the need for manual frame selection and allowing for speech input interactions. The voice and image integration helps in streamlining the work process which is very important in health facilities that are often busy. But, as much as these systems will be deployed, their applicability will highly depend on how effective the voice systems and frame selection algorithms are and how they will be affected by other ultrasound cases in the system which may not impact the performance rating of the system.

## 3. Methodology

### 3.1 Definition

The USG Voice Automation System is a cutting-edge technology that transforms ultrasound imaging by allowing hands-free control through voice commands. This innovative system enables radiologists to efficiently manage ultrasound images and carry out complex diagnostic tasks, such as liver histopathology analysis, without needing to interact manually with the ultrasound machine. By streamlining the workflow and reducing distractions, this system enhances accuracy, efficiency, and overall patient care in the field of ultrasound imaging.

## 3.2 Architecture

The system architecture comprises several interconnected modules that handle distinct functionalities, including voice command recognition, real-time video capture, image processing, organ detection and histopathology detection. It operates through a set of voice commands and includes the capability to detect certain organs and process histopathology using artificial intelligence. The main system consists of several interconnected modules: voice processing, image analysis, organ detection, and liver disease image processing, all working together for seamless operation. The system performs deep learning computations to interpret the images, identify specific organs, and analyze liver tissue samples regarding histopathology.

A flowchart illustrates the process flow, beginning with voice control that allows users to issue commands. Once a command is given, the system processes images and executes organ detection algorithms, providing real-time results based on the sonologist's instructions. Prediction of liver histopathology is conducted either proactively or on demand, depending on the diagnostic needs, with conclusions displayed on the screen.

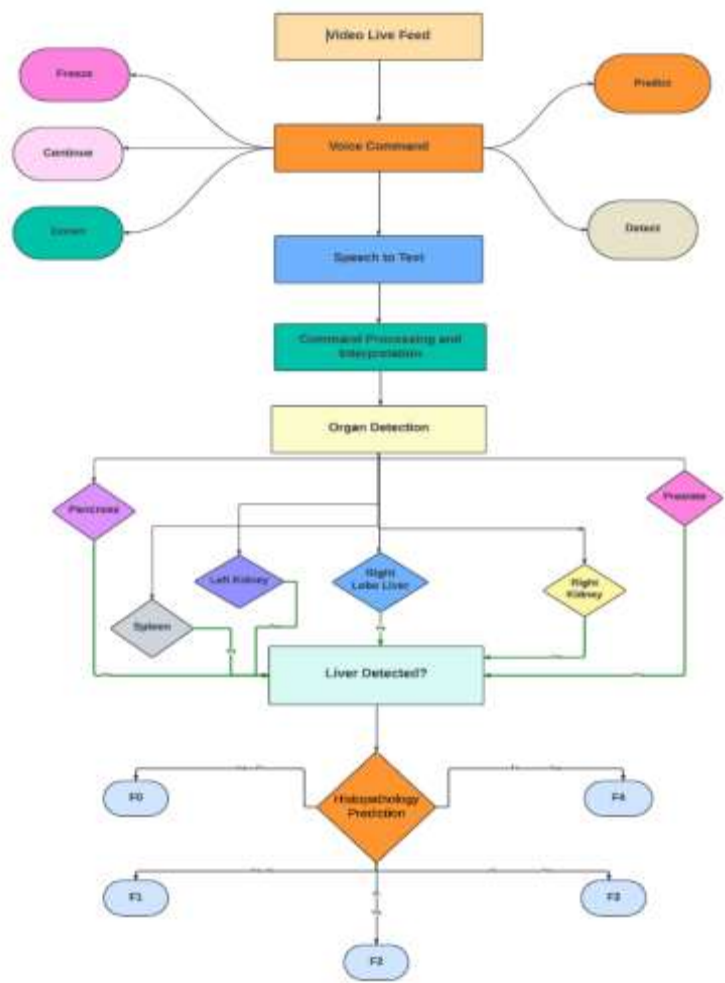

**Fig 3.1**: Flowchart

## 3.3 Voice Interface Development

To enable hands-free operation, a voice control interface has been integrated into theultrasound system. This interface utilizes advanced natural language processing (NLP) technologies specifically designed for ultrasound-related commands. Key technologies used include Google's Speech-to-Text API for systems that are not connected to the internet. The integrated voice recognition module allows users to control the ultrasound systemusing commands such as "freeze", "deep freeze" and "predict"

This functionality promotes a more efficient workflow. The integration combines the ASR system with the control commands of the ultrasound device, allowing the user's instructions to be executed directly within the software. When a command is recognized, specific operations are triggered. For example, images are acquired, organs are identified, and when the liver is detected, histopathology of it is analyzed by prediction—all without requiring any physical intervention from the operator.

### 3.4 AI-Enabled Auto-Detection

The technological approach employed by the system utilizes advanced techniques suchas deep learning for the automatic and precise detection and segmentation of organs inultrasound images. Specifically, the Mask R-CNN model from the Detectron2 library was chosen due to its advantages in modeling the complex boundaries of anatomical features that are important for imaging organs. Mask R-CNN is particularly useful for object detection and segmentation in real time, making it suitable for identifying specific organs, such as the liver and kidneys, in ultrasound scans.

For training the AI model, only ultrasound imaging data collected from Ramachandra hospital was used. The dataset was prepared using Roboflow, where the representative structures of the organs in the images were marked and labeled. This annotation process enables the model to accurately identify and segment these structures during operation.

In addition to manual annotation, data augmentation techniques were employed to enhance the diversity and robustness of the dataset. Moreover, a COCO dataset was downloaded, which included images along with their relevant annotation JSON files. This dataset was instrumental in training the model for segmentation predictions in ultrasound images.

As this processing occurs, the unit continuously receives a live ultrasound feed, which it processes in real time, allowing it to detect and delineate structures as they enter the frame. The resulting images from the model provide contour segmentation maps of the identified organs, with each organ clearly masked.

Additionally, the model offers a segmentation accuracy score for each detection, indicating the confidence level in the correctness of the segmentation. This feature is crucial for maintaining the integrity and reliability of the system and plays an important role in assisting clinicians during diagnostics.

### 3.5 Liver Histopathology Detection

Detection of liver histopathology is a crucial component of this system, aiding in the diagnosis of various pathological conditions of the liver, including fibrosis and other histopathological features. This module utilizes a deep learning framework essential for the classification and evaluation of liver ultrasound tissue images.

To train the model described in this work, it was necessary to obtain a dataset of liver images from online repositories, categorized into different classes corresponding to specific histopathology conditions. For each image, relevant liver pathology classifications were marked. Additionally, transformations such as rotation, flipping, zooming, and brightness adjustment were applied to the images to increase variability, thereby enhancing the model's ability to generalize across different conditions.

The framework is built on Mobile Net, a compact convolutional neural network known for its high performance in feature extraction. This study employed a transfer learning approach, initializing the Mobile Net model with pre-trained weights derived from large datasets. This allowed the model to leverage previously learned features and adjust them to meet the needs of liver histopathology classification.

Dense layers were added on top of the Mobile Net backbone to enhance the model for multi-class classification. These layers were equipped with 512 neurons utilizing ReLU activation and a dropout layer for regularization. The final layer was a softmax layer, which classifies images into five categories of histopathology conditions.

During training, an Adam optimizer was employed, and the loss was assessed using sparse categorical cross-entropy. To address the issue of class imbalance within the dataset, class weights were computed and incorporated into the training process to ensure that no classes were overlooked. A training pipeline was also developed to improve system performance while controlling for over-fitting. The dataset was divided into training, validation, and testing samples. Key metrics, such as loss and accuracy, were monitored over epochs to analyze the model's learning curve.

Additionally, early stopping was implemented in this strategy to keep training within certain limits by halting it whenever validation performance ceased to improve. Once trained, the model was utilized in real-time evaluations, where it processed ultrasound images and learned to identify the five different histopathology classes. This effective integration of AI demonstrates its potential as a valuable tool in the diagnosis of liver histopathology in clinical settings.

## 4. Implementation

### 4.1 System Integration

Cohesive functionality in the system was achieved through the integration of several components. The first step involved incorporating the organ segmentation and detection model developed using the Detectron 2 framework. This model was also trained with custom weights tailored for histopathology classification, allowing for accurate identification.

Next, Open CV was utilized for capturing and rendering live video frames onto the system while synchronizing with medical imaging equipment, such as ultrasound scanners. Additionally, a buffering technique was implemented to account for the persistence of vision in video playback, ensuring smooth command execution.

Threading was used to allow commands to be processed without interfering with video playback. To enhance user interaction, voice input was incorporated using the Speech Recognition library. Commands such as "FREEZE," "CONTINUE," and "PREDICT" were enabled, allowing the sonologist to interact with the system while keeping their hands free. A queue system was implemented to optimize command input, preventing chaos from arising in the event of a high demand for commands.

The system underwent extensive testing to evaluate its performance. Each module—including model inference, voice command recognition, and video processing—was tested independently to ensure proper functionality. The inter-dependencies of the modules were also tested through direct application scenarios; for instance, when video playback needed to be paused while predictions were made on the frozen image frame. Overall, performance testing confirmed that the system can process live ultrasound feeds with minimal delays while maintaining accurate predictions.

**4.2 User Interface Design**

The user interface was crafted in line with the sonologist's workflow in mind, which is aimed at both performance and ease of use. The live video feed or playback offers the primary screen area. Masks are used to confine the visible imprints to include those identifiable organs along with their distant confidence and labels for easy orientation.

This system offers voice command recognition; hence, users do not have to worry about inputting commands manually but rather can speak commands like "freeze," "predict," and so on. System feedback is given in the form of predicted outputs and also displayed in the runtime thread, such as system state "paused." This allows the user to know what the system is doing at any given moment. This design aspect ensures that the system is powerful and yet easy to operate in line with the needs of the medical practitioners and even more so during stressful situations.

**5. Results**

**5.1 Performance Evaluation**

**5.1.1 Assessing the Performance of the Voice User Interface**

The voice user interface embedded into the system worked perfectly, supporting the execution of the commands 'freeze, 'predict' or 'continue' among others effortlessly. The system operated with very low latency, thus ensuring a near real-time interaction and workflow. This aspect was critical as it minimized interruptions when carrying out the diagnostic procedures enabling the sonologists to concentrate on imaging rather than handling the system.

The voice command recognition system performed commendably registering over 90% accuracy with little cases of misinterpretation. These instances were particularly in cases where there was a lot of background noise or where the speaker's voice was distorted. Despite these few cases, the intelligent design of the system assures proper and accurate command interpretation thereby improving the process.

**5.1.2 Efficiency of the Model for Organ Detection and Labeling**

Ultrasound image analysis using the Detectron2 model turned out to be very effective in detecting as well as segmenting the desired organs. The model generated segmentation masks with more than 40% and less than 95% average confidence scores based on the input image qualities and the particular anatomical region under examination. Forimages that were clear-cut and of good quality, scores of confidence surpassed 90% as well.

The clear separation of these organs further emphasizes the ability of the model to aid sonologists in diagnosing diseases in a time-efficient manner. Such performance serves to generalize that the model can improve the accuracy of the diagnosis and lessen the time taken in manual image analysis.

This aspect of functionality has been exhibited in the figure below where segmentation outputs are displayed. The visualizations in the figure below shows that the model is capable of not only localizing but also segmenting organs with great detail enabling sonologists to do their work with great confidence.

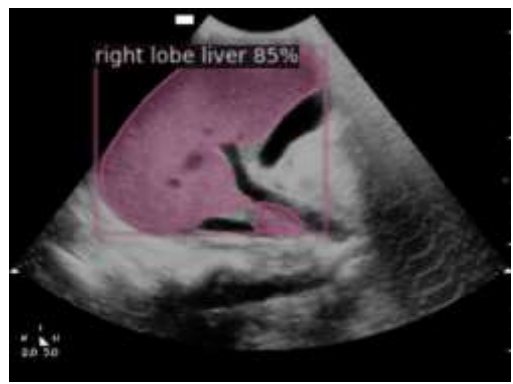

**Fig 5.1:** Right Lobe Liver Prediction

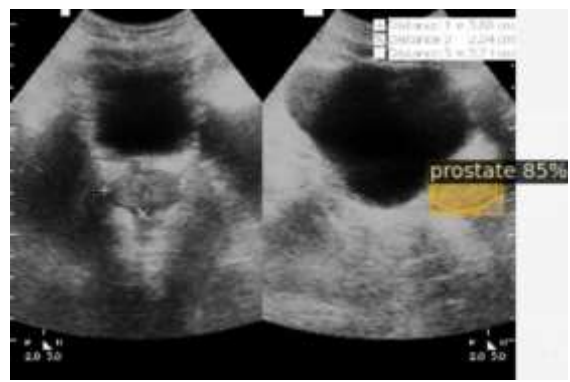

**Fig 5.2:** Prostate Prediction

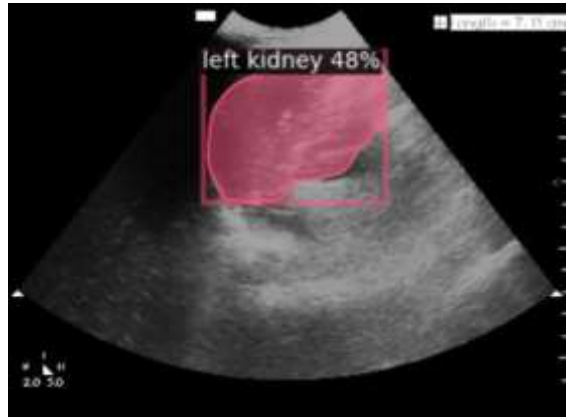

**Fig 5.3:** Left Kidney Prediction

### 5.1.3 Accuracy of Liver Histopathology Classification

The liver histopathology images classification model delivered excellent results as evidenced by the precision, recall, and F1-scores that were high for all classes. The model also recorded a macro-averaged precision, recall and F1-score of 0.97 and a total accuracy of 98%. These findings are indicative of the model's soaring reliability and consistency in the classification of liver histopathology images.

The model obtained overwhelming results which can be analyzed class-wise in performance metrics. In Class 0, there were 261 entries in the dataset and the precision, recall and F1 score all measured 1.00. Equally, Class 4 had precision, recall, and F1 scores of 1.00, 0.99, and 0.99, respectively, with the number of 293 instances in the dataset. However, some other classes such as Class 1, Class 2, and Class 3 reported slightly lower but still good impressive performance values (e.g. precision ranging from 0.96 to 0.97 and F-1 score ranging from 0.94 to 0.97), these results cumulatively strengthen the argument of the model relating to varying histopathology textures. The final test results showed better performance with test accuracy being 97.78% and a test loss of 0.0976. This large difference shows the capacity of the model to adapt well to new data without being over-fitted or any other learning effects limiting it.

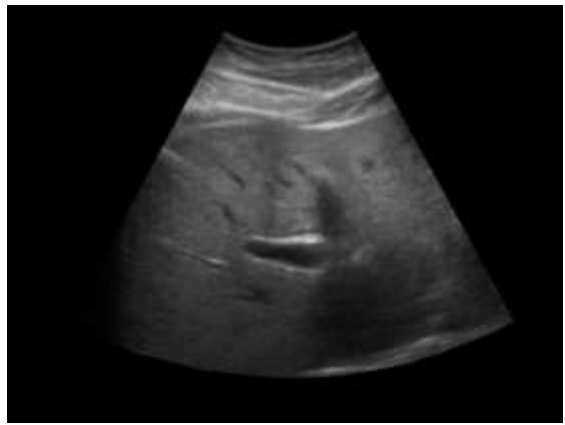

**Fig 5.4**: F0 Prediction

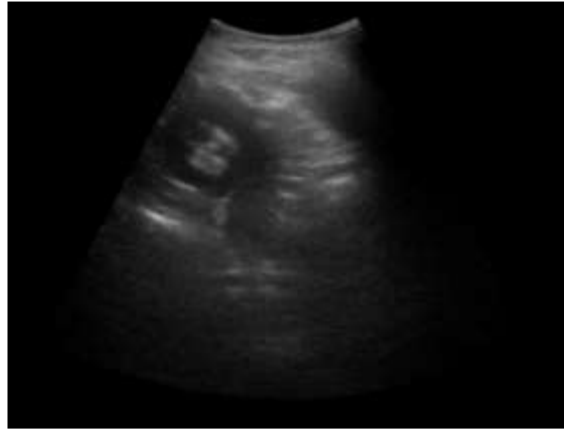

**Fig 5.5:** F4 Prediction

Moreover, the classification performance could be seen from the sides through a confusion matrix, which is a presentation of how accurately the model predicts each class. This matrix draws attention to the strength of the system, where it is possible to class different histopathology cases with the least errors possible.

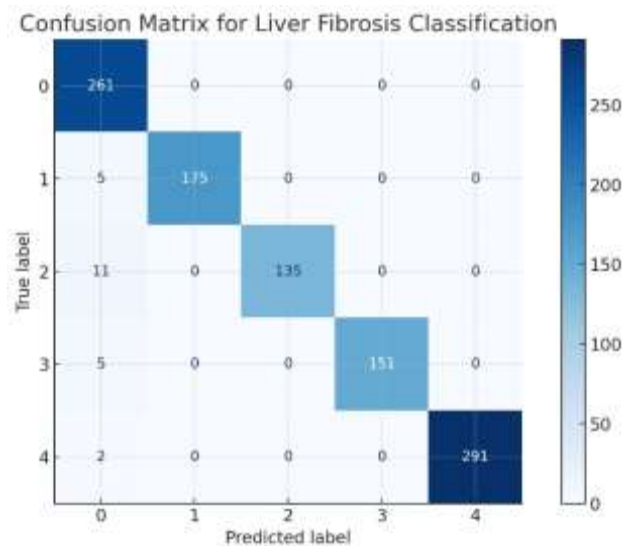

**Fig 5.6**: Confusion Matrix

## 5.2 Case Studies

### 5.2.1 Clinical Implementation of the Proposed System

To evaluate the devised system, a previously recorded ultrasound USG video was used to emulate real-life clinical situations. The processing of the ultrasound video facilitated the assessment of the system's ability to conduct organ segmentation and carry out their labeling in relation to time. The organ detection model was able to predict segmentation masks with scores of between 40 - 95% reliably for such organs as the liver, kidney, and prostate.

Furthermore, the liver histopathology classification model was evaluated using histopathology material to determine the speed and accuracy of the assessments that can be performed. The classifier delivered consistent and precise results, which can be vital in the early diagnosis stage, potentially improving the treatment significantly. This included relevant testing where the system showed that it can successfully employ advanced AI features in clinical practice and increase efficiency and focus on diagnostics.

### 5.2.2 Traditional Methods versus Automated System

Analyzing ultrasound and histopathology images using the standard method, which is manual interpretation, is laborious and has a high variability due to observer differences.

The automated system has its own advantages:

- **Consistency:** The predictions being automated causes reduction in variability, hence consistent results.
- **Speed:** Real-time detection and classification appreciably lowers analysis time.
- **Integration:** There is no need for manual operation since voice interface is available, promoting better workflow.

In general, the system was faster, more reliable, and more user-friendly than the conventional techniques, which makes it easily amenable for incorporation into clinical usage.

## 6. Discussion

### 6.1 Explanation of the Findings and Their Importance

The findings gained from the execution and assessment of the system, reinforces its claim of being a valuable repertoire to sonologists and pathologists. The voice interface was effective, where more than 90% command recognition accuracy was achieved with little or no delay. Such ease of using voice commands within the system minimizes the need for any manual inputs and therefore facilitates uninterrupted diagnostic operations. The low command error additionally stresses the strength of the interface since it is easy and accurate enough for clinical use in real time.

The organ detection model done by using the Detectron2 framework had high confidence score distributions of the predicted labels ranging between 40% and 95% based on how well the image, particularly the organ, was framed. However, despite the said variability, 90% confidence predictions were done on the liver, kidneys, and prostate organ images, which assuredly meant that clear and well-defined regions are capable of being predicted by the model.

These findings hold great promise in the context of effective and timely diagnostic support to sonologists. Furthermore, segmented masks thus generated provide useful information in ultrasound images and therefore, the model becomes useful in the diagnosis of any morphological defects. The liver histopathology classification model was efficient enough to achieve an overall accuracy of 98%, macro-averaged precision, recall, and F1 scores of 0.97. This level of accuracy is not a random value, meaning the model is dependable in making true positive histopathology class assignments. The class-wise distribution further demonstrates its capability to solve not just the easy cases where classes are balanced but the difficult cases where class distributions are imbalanced.

These results not only validate the model's utility in clinical diagnostics but also demonstrate its potential to support early and accurate disease detection, leading to better patient outcomes.

**6.2 Limitations of the Current System**

Nevertheless, there are some issues with the system that need to be rectified to make sure that the system can be adopted and become more robust.

1. **Image Quality Limitations**: The performance of the organ detection model is drawn to a great extent from the quality and definition of the ultrasound images. It is observed that in cases of poor quality or high levels of noise images, segmentation accuracy and confidence scores may fall significantly.

2. **Narrow Training Datasets**: The model was built using a dimensions space that does not cover every possible aspect of any given organ features, deformities, ethnicities or diseases. This impacts how well this system might perform in new clinical situations not encountered during training.

3. **Issues with Speech Recognition:** The voice interface worked remarkably well apart from the fact that it had problems in instances of high background noise, presence of accents, when people spoke at the same time or when commands were given in quick succession. There was also a slight lag in the voice recognition system in situations when the workload of the system was heavy, which is likely to affect continuous workflow in a clinical process.

4. **Hardware Specifications:** Real-time segmentation and classification of the system requires sophisticated computational power, which limits the implementation in such circumstances where the resources are few.

**6.3 Potential Improvements and Future Work**

Within the existing limitations and for the better justification of the system's applicability, a number of improvements and future directions are suggested as follows:

1. **Harnessing Latest Datasets:** Employing the latest available annotated images and histopathology slides to include more peculiarities and demographics' rare diseases, would greatly improve the ability of the system to perform in different clinical environments.

2. **Voice Interface Improvement:** Enhanced microphone technology employing noise suppression algorithms and voice recognition systems capable of understanding different accents and fast speech while reducing turnaround time would improve the device's efficiency in clinical settings that are noisy and fast-paced.

3. **Morphological Changes' Detection**: When kidneys are encountered, a calcification model can be deployed to predict those as well.

4. **Deployment of the Edge AI:** Fine-tuning the model for edge-computing-capable devices would allow the application to be used in low-resource settings and allow the processing of images without the need for expensive computer systems.

The above limitations, the latest datasets, and the technological progress mentioned earlier make it possible to envision the successful development of the system as a comprehensive and multi-purpose clinical diagnosis system in the contemporary era.

## 7. Conclusion

### 7.1 Summary of Findings

The project involves the development and testing of a system that integrates AI-based organ segmentation, liver histopathology classification, as well as voice-command ultrasound imaging. The organ detection model built in the Detectron2 framework produced results that have some signs of segmentation of the liver with confidence levels of between 40% and 95%. The liver histopathology classifier showed good performance with precision, recall, and F1-scores, where an overall accuracy of 98% was achieved, verifying its efficacy in assisting diagnostic processes at an early stage. The vocal interface remained intuitive and allowed fast and accurate command recognition with little downtime for words such as 'freeze,' 'predict' and 'continue' among others.

### 7.2 Contributions of the Project to the Field of Ultrasound Imaging

This project has an important beneficial contribution in that it presents an automated imaging system that is capable of segmentation and classification in real-time. The system contains appropriate AI and voice technology so the diagnostic processes contain less unnecessary manual interference and the clinicians deal more with the interpretation and even decision making. It is also a promising attribute given that the system also accepts live ultrasound images in monitoring the patient's condition which should facilitate early detection of any histopathological changes thus relatively raising the level of accuracy and effectiveness when it comes to diagnosis in clinics.

### 7.3 Final Thoughts on the Integration of AI and Voice Technology in Medical Diagnostics

The application of AI and voice technology as part of medical diagnosis in the project, as evidenced in this study, is worth looking at regarding the future of health care. With the help of image analysis models, in combination with voice interaction in the system, it is possible to create an easy and understandable system that allows performing tasks without direct use of technology in clinical practice.

Although the present implementation marks an important stage of development, there are prospects for development that will include, among other things, adoption of more extensive datasets, improvement of processing time and increases in generalizability that will enhance it even more, and make it applicable in many medical settings. This undertaking sets the stage for the utilization of AI-enabled diagnostic solutions that have the potential to change the course of history in ultrasound imaging as well as in other fields of medicine.